\documentclass[
    ,final            
  ]
  {aipproc}

\layoutstyle{6x9}

\begin{document}

\title{Searches for New Physics at Colliders}


\author{Giorgio Chiarelli}{
  address={INFN Sezione di Pisa, Via F.Buonarroti 2, I-56127 Pisa}
}



\begin{abstract}
In this paper I present the most recent results of the ongoing 
searches, mainly from Tevatron Collider experiments, for new physics 
beyond the Standard Model. 
While no signal has been seen so far, many analyses are reaching the 
point in which either a discovery will take place or strong limit on
currently popular theories will be set.
\end{abstract}

\maketitle


\section{Introduction}
The standard model of elementary particles and fundamental interactions 
(from now on Standard Model or SM) has been very successfull to describe 
the subnuclear world. Given this succes one 
would wonder why should we look at any theory beyond the standard 
model (BSM) at all.
There are several {\it theory oriented} answers to this question, from 
the hyerarchy problem to the stability of the Higgs sector, to the number 
of generations, just to name a few. 
In my view, however, most important are some hints of
open questions. What is the nature of the Cold Dark Matter (now that we 
have 
an indirect cosmological proof of its existence). Why neutrinos have 
masses? Is there a unification of the coupling constants before the Planck 
scale? Finally, while the SM has been succesful we still do not have a way 
to marry quantum theory with gravity, let alone to unify gravity with the 
other interactions.

An optimal place to look for new phenomena is the energy frontier which, 
at the moment, is explored by the experiments H1 and ZEUS at the HERA 
collider at DESY near Hamburg~\cite{hera}, 
where proton and electron (or positron) collide at $\sqrt{s}=320$ 
GeV/c$^2$ and at the Tevatron (experiments CDF and D0) where proton and 
antiproton collide at the highest energy (1.96 TeV) currently 
available~\cite{teva}.
In the near future that frontier will be throughly explored by the LHC, 
currently being built at CERN. Most of the results presented were 
obtained at the Tevatron where the goal of 1 fb$^{-1}$ delivered per 
experiment was exceeded in May 2005. Current scenarios predict between 4 
and 8 fb $^{-1}$ delivered (per experiment) by 2009.

This exploration physics proceeds mainly through comparison of data with 
SM expectations.
As {\it New Physics} (NP) is anything which is 
beyond the standard model, the best experimental way to 
look for it in a model independent way is to select
observables that can be affected by NP and then 
compare measurements with SM expectations. It takes a lot of 
ingenuity but it also implies that in those searches we will be studying 
familiar physics objects. Hadronic jets, charged and neutral leptons
are the basic tool.~\footnote{In the following we will use {\it MET} 
for Missing Transverse Energy.} 
Through this paper, unless otherwise indicated, 
95 \% C.L. limits will be reported.

\section{From the Standard Model to Beyond}
Although tested to a very good accuracy, the SM is incomplete as the Higgs 
particle, a cornerstone of the theory, has not been discovered.
The best limit for its mass, a free parameter of the theory, $M_H > 
114.4$ $GeV/c^2$, is due to the LEP 
experiments. At this mass the Higgs decays into two $b$ quarks most of the 
time.
At the moment the hunt for Higgs takes place at the Tevatron, where 
most of the production cross section is via $gg$ fusion ($\sigma \simeq 
0.8$ pb at the LEP limit). At low Higgs masses ($M_H \leq 130 \div 140 $ 
GeV/c$^2$) the signal is swamped by 
the much larger heavy flavour production. From an experimental point of 
view, at low mass, the associated 
production of vector boson ($W$ or $Z$) to Higgs ($\sigma 
\simeq 0.2 \div 0.1$ pb) has a clean signature and a better S/B ratio. 
 
\begin{figure}
  \includegraphics[height=.4\textheight]{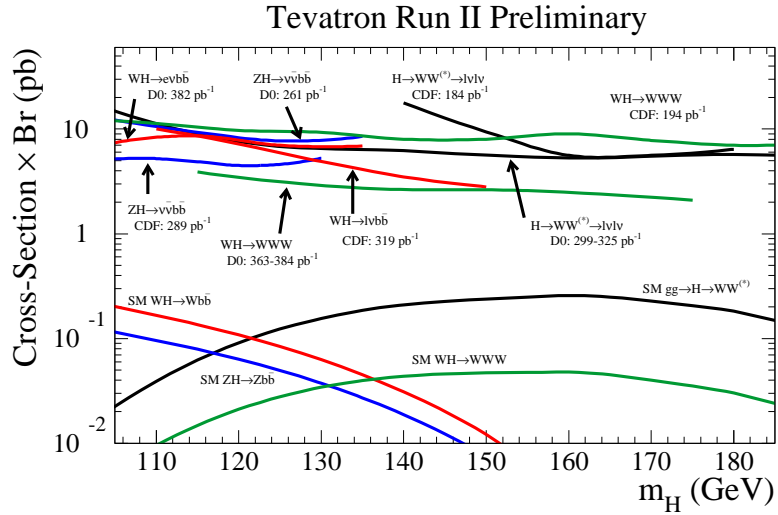}
  \caption{Summary of SM Higgs searches at the Tevatron. Summer 2005.}
\label{fig:tevhiggs}
\end{figure}

CDF and D0 searched for the SM Higgs in a large mass range, exploiting 
different channels. The negative searches can be summarized in a single 
plot (see fig.~\ref{fig:tevhiggs}) where recent results obtained with 
$300\div 400$ pb$^{-1}$ are 
shown~\cite{cdfd0}.

\section{SUSY Searches}
The scalar nature of the Higgs particle brings itself some technical 
problems to the SM. Its mass gets large corrections which must cancel out. 
Over the years this problem has been tackled in many ways, the most 
popular solution being the SuperSymmetry (SUSY). The idea, borrowed from 
the particle-antiparticle symmetry, is that there is shadow world of 
(super)particles. For any fermion a scalar partner would exist, while for 
each spin 0,1 particle a (s)fermion would exist~\cite{susy}. Clearly the 
symmetry is 
broken, as we do not have seen any sparticle. The nature of SUSY breaking 
is unknown, and there are several possible scenarios. However, all of them 
predict an enlarged Higgs sector (the single Higgs doublet of the SM is 
replaced by two). Also, in most models the assumption is that 
quantity $R=(-1)^{3(B-L)+2s}$ is conserved. That implies that in the final 
state the lightest supersymmetric particle (LSP) will 
escape detection leaving a typical $\nu$-like signature of missing energy. 
SUSY is strongly constrained by the (negative) findings of the LEP 
experiments. Masses of the sparticle are large therefore
currently, the best place to look 
for SUSY is the Tevatron where typical cross 
sections are $O(1)$ pb.
Due to the unknown 
nature of the symmetry breaking mechanism, the analyses are 
(mostly)aimed to specific channels/breaking mechanisms.
\begin{figure}
  \includegraphics[height=.35\textheight]{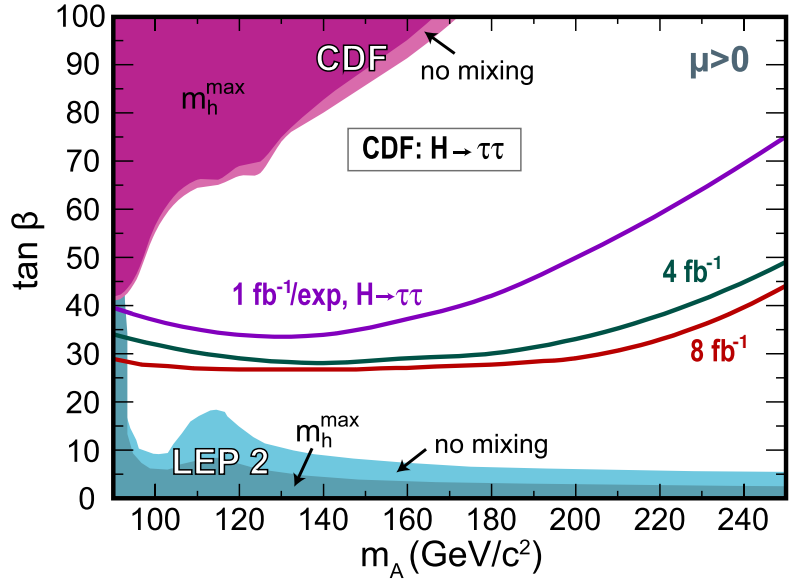}
  \caption{CDF limit for MSSM neutral Higgs $A$ (positive $\mu$ case).}
\label{fig:mssmh}
\end{figure}

\subsection{MSSM Higgs}
In the minimal supersymmetric model (MSSM), Higgs production can be 
enhanced, with respect to the SM case, for large $\tan \beta$ (where $\tan 
\beta$ is the ratio of the 
v.e.v of the two Higgs doublets).
There are five Higgs particles in the model (h, H, A and $H^{\pm}$), where 
$h$ and $H$ indicate a light (heavy) neutral Higgs and A a neutral 
pseudoscalar.
In this model the branching fraction for $H\rightarrow \tau \tau$ channel 
can be as large as $\simeq$ 8 \%. CDF developed a $\tau$ identification 
algorithm which yielded an $H\rightarrow \tau \tau$ global efficiency of 
about $0.5 \%$ for the $\tau_{l} \tau{h}$ decay channel where $l$ 
indicates an $e$ or $\mu$.
Exploiting its dataset of 310 pb$^{-1}$ CDF searched for the 
{\it h}(or {\it H}) and {\it A} particle (produced 
by $gg$ or $qq$) decaying into $\tau$ pairs. They found 236 events where 
263$\pm$ 30.1 were predicted~\cite{cdfmh}. The negative search is 
translated into a 
limit set within the framework of the MSSM (see fig.~\ref{fig:mssmh})).
With 1 fb$^{-1}$ or more, CDF expects to be able to explore the region 
down to $\tan \beta \simeq 
30\div 40$ for $m_{A} >120 GeV/c^2$ as shown in fig.~\ref{fig:mssmh}.

\subsection{Chargino and neutralino searches}
Exploiting the large production cross section for chargino and neutralino 
associated production, both experiments performed several searches.
The golden channel is the one in which there are three charged leptons in 
the final states, associated to the presence of large missing $E_{T}$ 
(signature of the LSP escaping detection). The low background and easy 
triggering on this channel is somewhat balanced by the need for good 
acceptance.
 
\begin{table}
\begin{tabular}{lcccccc}
\hline
  & \tablehead{1}{c}{b}{$e e l$}
  & \tablehead{1}{c}{b}{$\mu\mu l$}
  & \tablehead{1}{c}{b}{$\mu \mu $~\footnote{Same sign muons.}}
  & \tablehead{1}{c}{b}{$e\mu l$}
  & \tablehead{1}{c}{b}{$\mu \tau l$}
  & \tablehead{1}{c}{b}{$e \tau l$}\\
\hline
Expec.back.&$0.21\pm 0.12$&$1.75\pm 0.57$&$0.66\pm 0.37$&$0.31\pm 
0.13$&$0.36\pm 0.13$&$0.58\pm 0.14$\\
Obs.&$0$&$2$&$1$&$0$&$1$&$0$\\
\hline
\end{tabular}
\caption{D0 search for chargino in trilepton channels.$l$ indicates an 
high $P_T$ isolated track.}
\label{tab:d0tril}
\end{table}

\begin{figure}
  \includegraphics[height=.3\textheight]{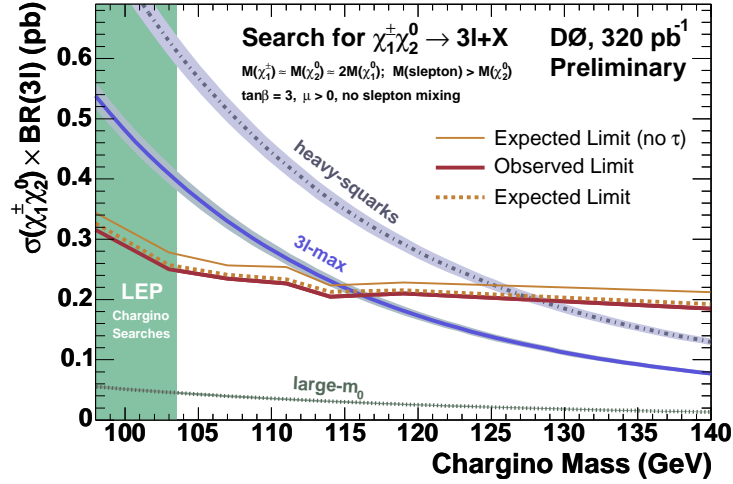}
  \caption{D0 limit for chargino mass from trileptons 
searches.}
\label{fig:charg}
\end{figure}
D0 searched into six different channels (see tab.~\ref{tab:d0tril}) in 320 
pb$^{-1}$. SM background is different from channel to channel, mostly due 
to misendified leptons and to diboson production. The 
negative result (4 events observed where 3.85$\pm$0.75 are expected) 
is converted into limits in the
framework of a specific mSUGRA scenario (light sleptons, low $m_0$) and 
yield a limit for 
$M_{\chi^{\pm}} >117$ 
GeV/c$^2$ (see fig.~\ref{fig:charg})

\begin{figure}
  \includegraphics[height=.4\textheight]{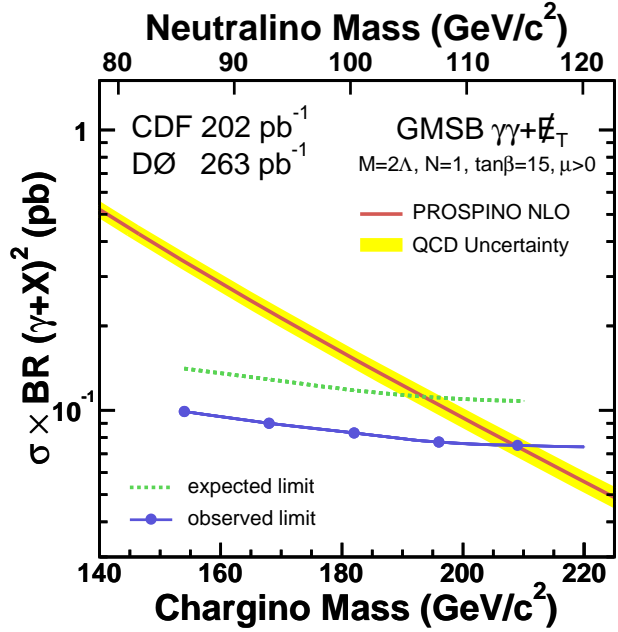}
  \caption{Combined CDF/D0 chargino mass limit in GMSB model.}
\label{fig:gmsbc}
\end{figure}

In Gauge Mediated SUSY Breaking (GMSB) models, gravitino becomes the 
LSP and neutralino becomes the next LSP (NSLP). Therefore neutralino 
decays 
into gravitino-$\gamma$. For short neutralino lifetimes the experimental
signature is a final state with 
2 high $E_T$ $\gamma$ and MET. 
CDF searched 200 pb$^{-1}$ for events with MET$>45$ GeV and 2 $\gamma$s 
with 
$E_T>13 $GeV, D0 looked into 263 pb$^{-1}$ for events with 
MET$>40$ GeV and 2$\gamma$ with $E_T>20$ GeV. Both searches gave a 
negative result, and the combined limit is set at a chargino mass of 209 
GeV/c$^2$ (see fig.~\ref{fig:gmsbc})~\cite{cgmsb}.

\subsection{Gluino and Squarks}
\begin{figure}
  \includegraphics[height=.4\textheight]{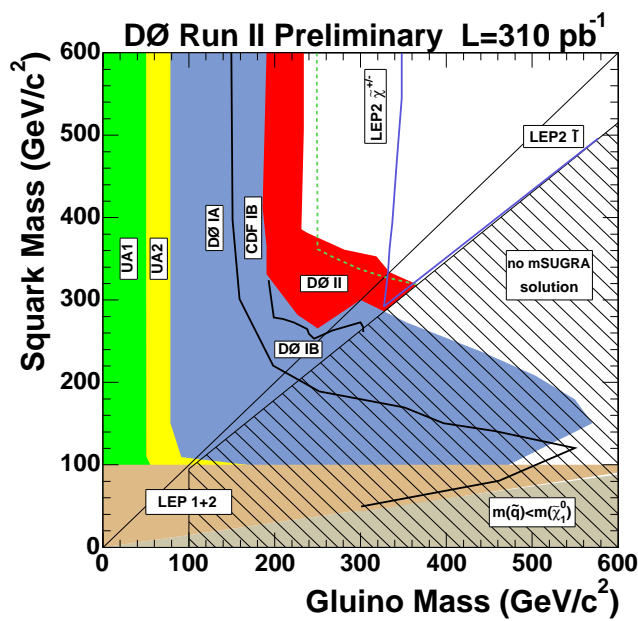}
  \caption{Gluino-squark mass limit form D0 searches}
\label{fig:d0gq}
\end{figure}

Decays of $\tilde{g}$ and $\tilde{q}$ at the Tevatron produce 
multijet events with 
missing energy. 
Both experiments looked 
for events with such topologies.
D0 search for $\tilde{g}$ and $\tilde{q}$ was optimized as 
different 
regions of the $\tilde{g}-\tilde{q}$ mass plane were explored 
by using three different subsamples divided according to the number of 
jets and using $H_T$ cut to reduce multijet background. Di-jet events 
($H_T>250$ GeV and MET$>$175 GeV) were selected to search in the region 
$m_{\tilde{g}}\geq m_{\tilde{q}}$. 
Three jet events with $H_T>325$ GeV and MET$>$100 GeV 
to cover the region $m_{\tilde{q}}\sim m_{\tilde{g}}$ and finally four 
jet events ($H_T>250$ GeV, MET$>75$ GeV) to cover the region
$m_{\tilde{q}}\gg m_{\tilde{g}}$. After optimizing the selection to reduce 
multijet background, the final results are the following:12 events 
found where $12.8\pm 5.4$ are expected in dijet sample, 5 events found in 
three jet sample where $6.1\pm 3.1$ are expected and 10 events found where 
$7.1\pm0.9$ are expected in four jet sample.
Limits are then set 
within the mSUGRA model for the scenario: $\tan \beta =3$, $A_0=0$, $\mu 
<0$ (see fig.~\ref{fig:d0gq}). 

\begin{figure}
  \includegraphics[height=.4\textheight]{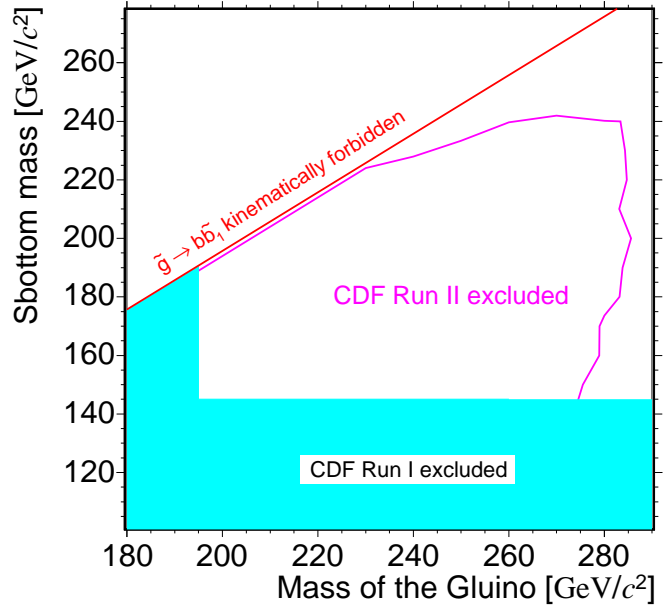}
  \caption{CDF limit on $\tilde{g}\rightarrow \tilde{b}$ decays.}
\label{fig:cdfsb}
\end{figure}

If third generations $\tilde{q}$ are involved, specific 
final states with heavy flavours can be favoured. Exploiting samples 
enriched in $b$-jet, 
limits on the scalar partner of the $b$ quark ($\tilde{b}$) were set by 
both experiments.
CDF searched for $\tilde{g}\rightarrow \tilde{b}$ decays with $\tilde{b} 
\rightarrow {b}+LSP$ in 156 pb$^{-1}$ of data. 
The search, performed looking for a final state with MET and four jets 
(two of them positively identified as containing $b$ quark debris), gave a 
negative result. The corresponding exclusion plot is shown in 
fig.~\ref{fig:cdfsb}.
D0 searched for signals of direct production of sbottom pairs decaying 
into $b$ and neutralinos. Using b-tagging to improve S/B ratio they are 
left (in 310 pb$^{-1}$) with events consistent with SM expectations and 
excluded sbottom masses up to $\simeq 200$ GeV/c$^2$ for neutralino masses 
up to 80 GeV/c$^2$.

\subsection{SUSY indirect searches}

A very important way to test for SUSY is to look at effects on SM 
processes due to SUSY contributions. A promising channel is 
$B_{s,d} \rightarrow \mu \mu$ as $B_s$ (and $B_d$) are copiously produced 
at 
the Tevatron. SM branching fraction (BF) is $(3.5 \pm 0.9)\times 10^{-9}$ 
but, in SUSY 
models, 
can be enhanced by several order of magnitudes for large $\tan \beta$.
CDF and D0 searched in 364 and 300 pb$^{-1}$ of data.
CDF set 90(95) \% C.L. limit to $BF(B_s \rightarrow \mu \mu) <1.5 (2) 
\times 10^{-7}$ and $BF(B_d \rightarrow \mu \mu) <3.8 (4.9)\times 10^{-8}$ 
(better than 
BaBar limit of $8\times 10^{-8}$). D0 limit at 90 \% C.L. is $BF(B_s 
\rightarrow \mu \mu) <4.1 \times 10^{-7}$ which, combined with CDF result, 
yield a Tevatron limit of $1.2 \times 10^{-7}$ at 90 \% C.L. 
Those results set a strong limit on MSSM inspired SUSY models~\cite{www}. 

\subsection{RPV SUSY}
\begin{figure}
  \includegraphics[height=.35\textheight]{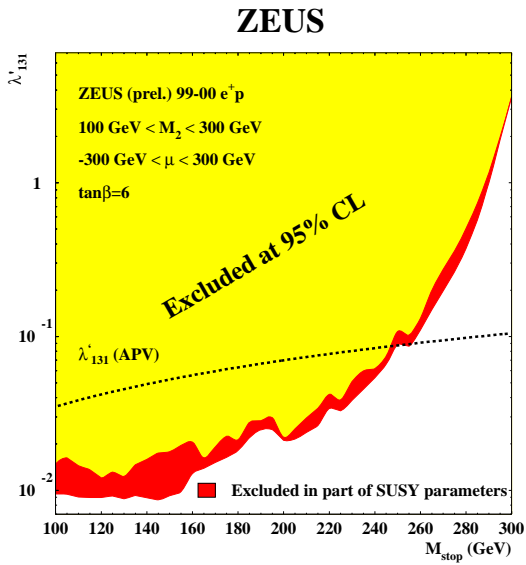}
  \caption{ZEUS limit on stop mass in RPV SUSY.}
  \label{fig:zeustop}
\end{figure}

In some SUSY theories quantity $R$ is not 
conserved, that implies that MET (due to escaping LSP) is not a typical 
signature. The relevant parameter is represented by the couplings among 
generations.
Searches for RPV SUSY take place at both the Tevatron and HERA. At the 
Tevatron CDF searched, without success, for sneutrino production and 
set limit to $m_{\tilde{\nu}}>$ 205 GeV/c$^2$ for the coupling 
$\lambda_{311}=0.01$.
ZEUS and H1 set limits to $\tilde{t}$ production. In 
fig.~\ref{fig:zeustop} you see that for $\lambda\simeq 0.01$ 
$M_{\tilde{t}}>260$ GeV/c$^2$ at 95 \% C.L.

\section{Living on a Brane}
In recent years, a new paradigm for the physics Beyond the Standard Model 
emerged. It is linked to the discovery that, given current theoretical and 
experimental constraints, there is room for large extra-dimensions. The 
basic idea is that particles are not 
point-like objects, but rather extended strings. The main interest of 
this class of theories is 
that they look like a very promising way to quantize gravity. Indeed the 
weakness of gravitation in our world is explained as being the remnant (in 
standard 3+1 D world) of a stronger interaction which propagates into the 
extra (hidden) dimensions which are then compactified. 
There are several options for the mechanism to work. 
In the Arkani-Hamed, Dimopolous, Dvali (ADD) theory 
the relevant parameter is the string scale $M_s$ which relates the Planck 
scale to the compatictification radius. In Randall-Sundrum theories the 
gravity is localized in the Extra Dimensions (which are highly warped) and 
the relevant parameters are the compactification radius and the curvature 
scale thate relate the Planck scale to the TeV scale.
As each particle is seen as part of a tower of Kaluza-Klein excitations, 
the main phenomenological effect is the existence of such states. At the 
Tevatron both direct searches for gravitons (RS model) as well 
as indirect searches of effects on SM processes (ADD model) were 
performed.
In the latter category, 
CDF and D0 searched for modification to $ee$, $\mu \mu$, $\gamma \gamma$ 
production due to graviton exchange. 
D0 searched in its data by studying its double differential cross 
section:
$${{d^2 \sigma}
\over{dMd\cos\theta^*}}=f_{SM}+f_{int}\eta_G+f_{KK}{\eta_G}^2$$ 
The effect of the 
existence of extra-dimensions is parametrized (using $f$ and 
$\eta_G$) according to different 
theoretical assumption. Data is consistent with SM expectations and limits 
are set in different scenarions. In table~\ref{tab:d0extrad} we show
a summary of the results in various channels according to several 
theoretical options (Giudice-Rattazzi-Wells, Hewett, Han-Lykken-Zhang)
\begin{table}
\begin{tabular}{lccccc}
\hline
  & \tablehead{1}{c}{b}{GRW}
  & \tablehead{1}{c}{b}{Hewett}
  & \tablehead{1}{c}{b}{HLZ, n=2}
  & \tablehead{1}{c}{b}{HLZ, n=3}
  & \tablehead{1}{c}{b}{HLZ, n=4}\\
\hline
$ee + \gamma \gamma$&$1.43$&$1.28$&$1.67$&$1.70$&$1.43$\\
$\mu \mu$&$1.09$&$0.97$&$1.00$&$1.29$&$1.09$\\
\hline
\end{tabular}
\caption{D0 combined (RunI-II) limits for large extra dimensions.}
\label{tab:d0extrad}
\end{table}

\begin{figure}
  \includegraphics[height=.3\textheight]{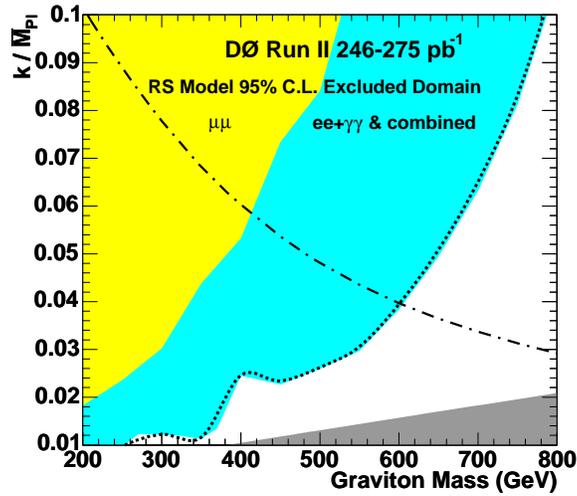}
  \caption{D0 limit on the mass of R.-S. graviton.}
  \label{fig:d0rsg}
\end{figure}

In direct searches, both experiments looked for
signal of direct production of the Randall-Sundrum graviton (a spin 2 
particle) in $\gamma \gamma$ and dilepton ($e$ and $\mu$) 
samples. CDF analized 345 pb$^{-1}$ of data and D0 looked at 
$\simeq 250$ pb$^{-1}$ of data. Results are 
consistent with SM expectations and limits are set. The 
strongest constraint is obtained by D0 that, for $k/M_{Pl}=0.1(0.01)$ set 
a 95 \% C.L. limit of $M_G>785 (250)$ GeV/c$^2$
(fig.~\ref{fig:d0rsg}).

\section{Searching in the dilepton sample}

The dilepton channels have been through investigated by both experiments 
as any deviation from SM would easily signal new physics. 
The measured quantity is $\sigma \times BF$, and negative findings can be 
interpreted as limits for physical states by selecting a specific 
theoretical model. 
Indeed, while the 
acceptance has strong dependence mainly from the spin of the mother 
particle, an important role is played by the (unknown) couplings. 
\begin{figure}
  \includegraphics[height=.3\textheight]{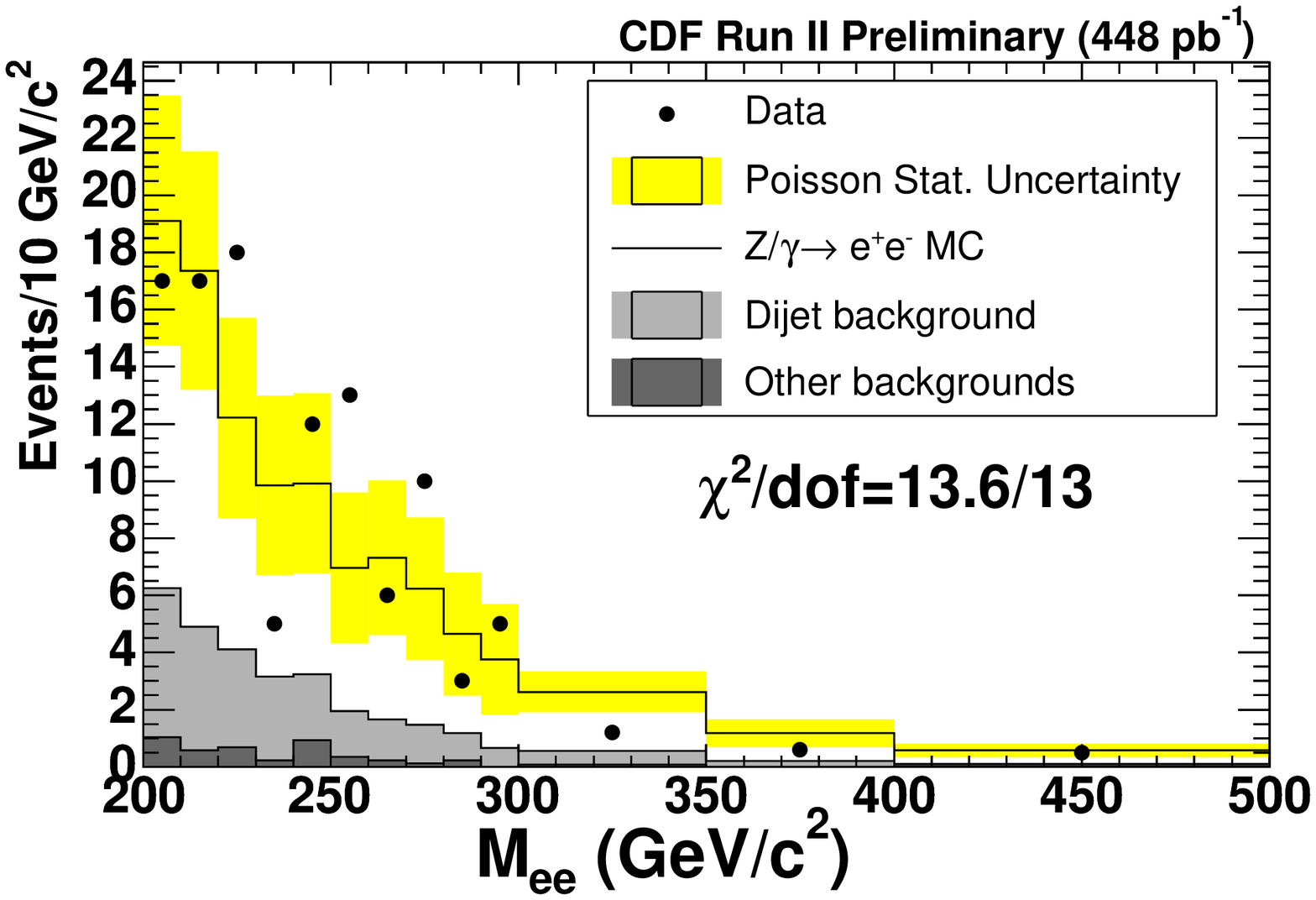}
  \caption{CDF Zp}
  \label{fig:cdfzp}
\end{figure}
CDF recentlty presented a preliminary analysis of data in 
the $ee$ channel. In fig.~\ref{fig:cdfzp} the search region ($M_{ee}> 200$ 
GeV/c$^2$) is shown. No visible excess over SM expectations is seen in 
448 pb$^{-1}$.
In order to convert 
negative findings into limits, model-specific calculations are performed. 
Also, a standard way to estimate the sensitivity to a $Z$-like particle is 
to assume SM couplings. The CDF result in 448 pb$^{-1}$ 
converts into a limit on SM-like $Z'$, $M_{Z'}> 845$ at 95 \% C.L.
In table~\ref{tab:zprim} we show the summary
of searches in different decay channels. 
\begin{table}
\begin{tabular}{lccccccccc}
\hline
  & \tablehead{1}{c}{b}{SM($ee$)}
  & \tablehead{1}{c}{b}{SM($\mu \mu$)}
  & \tablehead{1}{c}{b}{SM($\tau \tau$)}
  & \tablehead{1}{c}{b}{$Z_I$}
  & \tablehead{1}{c}{b}{$Z_{\chi}$}
  & \tablehead{1}{c}{b}{$Z_{\psi}$}
  & \tablehead{1}{c}{b}{$Z_{\eta}$}\\
\hline
CDF&845&735&394&610&670&690&715\\
D0&780&680&-&575&640&650&680\\
\hline
\end{tabular}
\caption{Limit for a sequential $Z$. CDF limits for $E_6$ $Z'$ refer to 
$ee$ and $\mu \mu$ channel combined using 200 pb$^{-1}$.}
\label{tab:zprim}
\end{table}

D0 also analized its dilepton samples ($ee$ and $\mu \mu$) looking for 
effects due to {\it preons}, the putative elementary constituents of 
quarks and leptons. The effects are parametrized in terms of $\Lambda$, 
the compositness scale. As for the LED case, the search compared the 
spectrum invariant mass and scattering angle of data with SM expectations.
The parametrization used is:
$${{d^2 \sigma} 
\over{dMd\cos\theta^*}}=f_{SM}+{I\over{\Lambda}^2}+{C\over{\Lambda}^4}$$ 
where $I$ and $C$ are input from theory and express the possibility of 
constructive/destructive interference. 
The negative result is used to set limits on the fundamental scale 
$\Lambda$. Limit is 
$\Lambda> 
3.6(9.1)$ TeV for constructive(destructive) interference in the 
electron channel in 271 pb$^{-1}$. In the muon channel, using 400 
pb$^{-1}$ the 
limit is stronger and set to
$\Lambda>4.2(9.8)$TeV for constructive(destructive) interference.

\section{Leptoquarks}
The fermionic sector of the SM can be symmetrized by the introduction of 
still unknown particles -LeptoQuarks (LQ)- carrying both barionic and 
leptonic numbers and coupling to both quarks and leptons. Relevant 
parameters for the theory are the couplings, $\lambda_{ij}$ and the decays 
branching fractions, which are ordered according to the BF into charged 
leptons ($\beta$) which can vary between 1 and 0. 

\begin{figure}
  \includegraphics[height=.4\textheight]{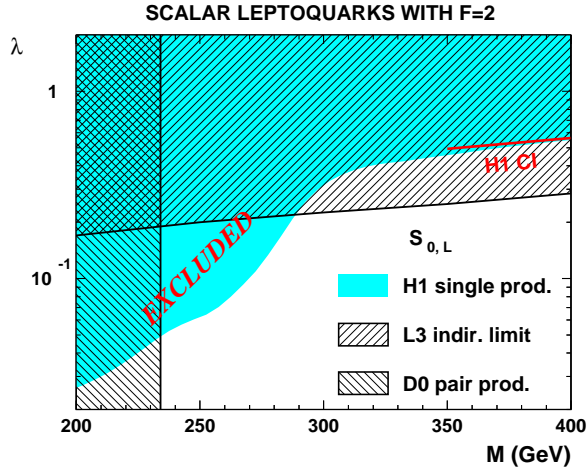}
  \caption{H1 limit for LQ production.}
  \label{fig:h1lq}
\end{figure}

At HERA first generation LQs can be produced in resonant process.
Both H1 
and ZEUS searched for signals of LQ production in their CC 
and NC samples. No signal is found and limits are set. Due to the way LQs 
are produced those limits are a function of the LQ mass and of the 
couplings. In fig.~\ref{fig:h1lq} we show the result from H1 
which, for 
a coupling strength $\simeq 0.3$ is $M_{LQ}>290$ GeV/c$^2$.

\begin{figure}
  \includegraphics[height=.28\textheight]{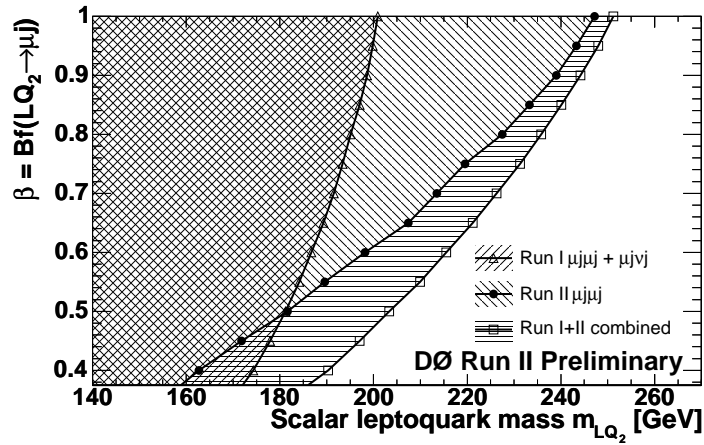}
  \caption{D0 limist for second generation LQ.}
  \label{fig:d0slq}
\end{figure}

At the Tevatron both experiments searched for scalar LQs in all 
generations.
As LQs are pair-produced, the actual measurement can be one of the 
following: $\sigma \times \beta^2$, 
$2\times \sigma \beta (1-\beta)$,$\sigma\times (1-\beta)^2$, depending 
upon 
the BF 
into charged leptons. The latter channel has a final state of two jets and 
MET and the analysis performed is generation-independent. CDF report a 
negative result in this channel and set a limit, using 
191 pb$^{-1}$, to $M_{LQ}>117$ GeV/c$^2$. More interesting are the 
results for 
first and second generation LQ. In those cases both experiments analized 
their samples $ll jj$, $l \nu jj$, $\nu \nu jj$ where $l=e$,$\mu$.
CDF reports a negative result and set 
limits, for $\beta=1$ to $M_{LQ}> 235(224)$ GeV/c$^2$ for 1$^{st}$ 
(2$^{nd}$) generation LQ in 200 pb$^{-1}$. 
\begin{figure}
  \includegraphics[height=.3\textheight]{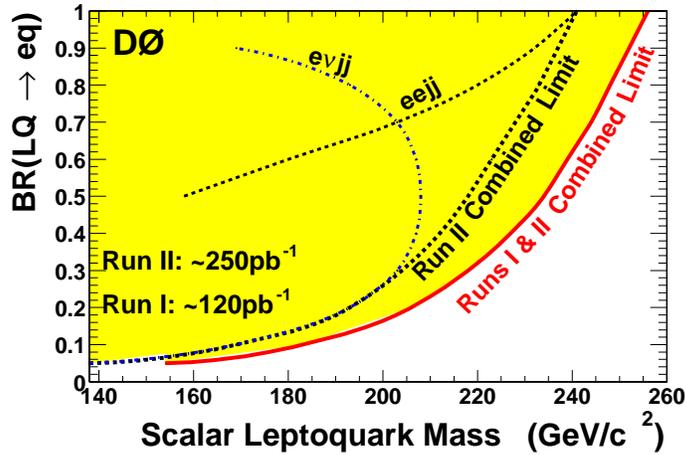}
  \caption{D0 limits for first generation LQ.}
  \label{fig:d0flq}
\end{figure}
D0 
combines its negative results in 250 pb$^{-1}$ (1$^{st}$) and 
294 pb$^{-1}$ (2$^{nd}$) generation with Run I results and set limits for 
$\beta=1$ to $M_{LQ}>256(251)$ GeV/c$^2$ for 1$^{st}$(2$^{nd}$) generation 
LQ (see fig.~\ref{fig:d0flq},~\ref{fig:d0slq}).

CDF searched for direct production of 3${rd}$ generation LQ in an analysis 
which also set limits on RPV SUSY stop production. As $4.8\pm 0.7$ 
events are found in 200 pb$^{-1}$, where 5 are expected a limit is set 
$M_{LQ}>$ 129 GeV/c$^2$.

\section{Conclusion}
A number of searches for signals of physics beyond the SM are ongoing. 
While their current findings do not support evidence of such processes, a 
very exciting period is ahead of us. The expectations of CDF and D0 to
collect 4 to 8 fb$^{-1}$ per experiment by 2009, as well as the LHC startup, 
indicate that within a few years either we will have 
seen signal of non-SM physics, or a number of theories will be disproved.


\begin{theacknowledgments}
I would like to thanks my colleagues from the Tevatron and HERA for the 
preparation of this talk, and in particular B.Heinemann, S.Ming Wang, 
J.F.Grivaz, M.Wing, E.Tassi, E.Gallo, A.Anastassov.
\end{theacknowledgments}



\bibliographystyle{aipproc}   





\end{document}